\documentstyle[12pt,fleqn,cite,epsfig]{article}
\newcommand{\sect}[1]{\setcounter{equation}{0}\section{#1}}

\def\ab {(a)}

\def\ten{(10)}
\def\be{\begin{equation}}
\def\ee{\end{equation}}
\def\bea{\begin{eqnarray}}
\def\eea{\end{eqnarray}}

\begin{document}
\begin{titlepage}
\begin{center}
\hfill hep-th/0112219\\
\hfill IP/BBSR/2001-26\\
\vskip .2in

{\Large \bf D-brane Bound States from Charged Macroscopic Strings}
\vskip .5in

{\bf Alok Kumar\footnote{e-mail: kumar@iopb.res.in}, Sudipta
Mukherji\footnote{e-mail: mukherji@iopb.res.in} and Kamal L. 
Panigrahi\footnote{e-mail: kamal@iopb.res.in}\\
\vskip .1in
{\em Institute of Physics,\\
Bhubaneswar 751 005, INDIA}}

\end{center}

\begin{center} {\bf ABSTRACT}
\end{center}
\begin{quotation}\noindent
\baselineskip 10pt
We construct new D-brane bound states using
charged macroscopic type IIB string solutions. 
A generic bound state solution, when dimensionally reduced, carries
multiple gauge charges. Starting with 
$D=9$ charged macroscopic strings, we obtain solutions in
$D=10$, which are interpreted as carrying ($F$, $D0$, $D2$) charges as well
as nonzero momenta. The masses and charges are also explicitly shown to 
satisfy the non-threshold bound of $1/2$ BPS objects. 
Our solutions reduce to the known D-brane bound state
solutions with appropriate restrictions in the parameter space.
We further generalize the results to  $(Dp- D(p+2))$ bound state
in IIA/B theories, giving an explicit example with $p=1$.

\end{quotation}
\vskip 2in
December 2001\\
\end{titlepage}
\vfill
\eject
\sect {\bf Introduction}

Bound states of D-branes 
\cite{witten,mli,mrd,green,russo-tsey,costa,araf,jabb} 
have been an interesting area of investigation due to 
their applications in understanding the non- perturbative aspects 
of various string and gauge theories \cite{malda-russo,hashi,oz}. 
In particular, the supergravity configurations of  supersymmetric D-branes
and their bound states were used extensively in testing various 
conjectures which involve knowledge of string theory or gauge
theory beyond their perturbative regime. 
In view of their widespread
applications, it is often very useful to generalize D-brane bound
state constructions.
As a step in this direction, in the present paper, 
we first give a construction of generalized $D0 - D2$ bound state using
charged macroscopic strings \cite{sen92}. These strings carry, in general,
many parameters associated with the charges and currents. As a consequence,
the D-brane bound states have many nontrivial charges. We argue
that these  configurations can also in general carry 
F-string charges and momenta along their worldvolume directions. 
We then further extend our result by 
presenting the generalized $(Dp-D(p+2))$ bound states. In this context
we give an explicit example for the $D1-D3$ case.  


A method for obtaining the bound state of various D-branes is
described explicitly in \cite{myers}. This is done  by smearing the brane 
of type IIA (IIB) theory along certain transverse direction
and mixing it with  a longitudinal direction through a 
coordinate transformation to construct `tilted' brane in $D=10$. Finally,
an application of $T$-duality on the configuration leads to a bound state
solution in the dualized IIB (IIA) theory. As stated earlier,  
in this paper we concentrate mainly on $D0-D2$ bound state
of the IIA theory.     
They are constructed by starting with a $D$-string solution in 
the IIB theory in $D=10$, which is also delocalized along
one of the transverse directions. These 
delocalized solutions can also be obtained by 
decompactifying a fundamental string solution in $D=9$ to $D=10$.
Further, one applies an
$S$- duality transformation in $D=10$ to construct $SL(2, Z)$ multiplets.
One, however, knows the existence of more 
general string solutions in heterotic as well as type II strings in 
all dimensions $4 \leq D \leq 9$\cite{sen92,kumar}, 
which carry vectorial 
charges and currents. We make use of such charged macroscopic string solutions
to generate  a generalized smeared (delocalized) $D$-string in $D=10$. 
We then apply the procedure for obtaining the D-brane bound states,
as described above, on such delocalized configurations. 
For the special case
when vectorial charges are set to zero, our delocalized  
solutions reduce to the $D=10$ smeared string solutions of  
\cite{myers}. We therefore have a generalization of the $D0 - D2$
bound state by starting with the charge macroscopic string 
solutions in $D=9$. 

We also show that by smearing some of the transverse directions of our
`generalized' $D0-D2$ bound state discussed above, and applying 
$T$-duality along these additional directions, we can construct 
new $Dp-D(p+2)$ bound states as well. We work out the 
case, $p=1$, explicitly. 
In fact even more general bound states can be constructed 
by using $D<9$ charged macroscopic 
string solutions. For the purpose of this paper, we however 
mainly restrict ourselves to the $D=9$ solutions. 
We also explicitly
verify the (non-threshold) BPS condition in all the examples 
discussed in the paper.

The charged string solutions are  
generated from the neutral ones by a solution generating transformation. 
They are parametrized by a group $O(d-1, 1; d-1,1)$, where $d$ is the
number of compactified directions. 
In particular, the solution described in \cite{sen92,kumar} 
is parametrized by two nontrivial parameters of the transformation 
namely $\alpha$ and $\beta$. The general solution for arbitrary values
of $\alpha$ and $\beta$ for $D < 10$ is given in \cite{kumar}.
Explicit supersymmetry property of these solutions are given 
for $ \beta=0$, $\alpha\ne0$ and $\alpha = -\beta$ in \cite{kumar}. 
For algebraic simplifications, in this paper, we will deal 
with these choices  
of $\alpha$ and $\beta$. Decompactifications, $SL(2, Z)$ transformations
and $T$- duality operations for general $\alpha$ and $\beta$ are 
possible to write down, but algebraically more complicated.    
The rest of the paper is organized as
follows. In section-2, we present a review of the charged macroscopic
string solutions.  In section-3, we first review the construction 
of  $(D0- D2)$ bound state in $D=10$, starting from $D$-string
solutions. We then discuss the construction of $Dp-D(p+2)$ bound
states from these ones by application of $T$-duality along 
other (smeared) transverse directions. 
Generalization of these non-threshold bound states, 
replacing the neutral strings by charged macroscopic strings
is presented for the specific solution $\alpha = -\beta$ in section-4.
Results for $\beta = 0, \alpha \neq 0$ solution are presented in
section-5. Here we also present further generalization of the results 
by using $(p, q)$ type IIB strings and 
show the existence of the non-threshold bound states for any
$(p, q)$-string, generated from a D- string via $SL(2, Z)$. 
We explicitly calculate the ADM mass and verify
the mass and charge relationship in these examples 
to check BPS nature of these configurations.
Conclusions and discussions are presented in section-6.

\sect{\bf Charged Macroscopic Strings}
\subsection{Charged Macroscopic String Solution}
   
We start by writing down the bosonic backgrounds associated with 
the charged macroscopic strings in space-time dimensions $D$ for type II 
theories. For charged string solutions below, 
we use notations and conventions that are identical to the
ones in \cite{sen95} and \cite{kumar}. The solution is given by,
\bea 
ds^2 &=& r^{D-4} \Delta^{-1} [ -(r^{D-4}+C) dt^2 + C (\cosh\alpha -
\cosh\beta) dt dx^{D-1} \cr
& \cr
&+& (r^{D-4} + C \cosh\alpha \cosh\beta) (dx^{D-1})^2]\cr
& \cr
&+& (dr^2 + r^2 d\Omega_{D-3}^2),     
\label{e13}
\eea
\be \label{e14}
B_{ (D-1)t} = {C\over 2\Delta} (\cosh\alpha + \cosh\beta)
\{ r^{D-4} + {1\over 2} C (1 + \cosh\alpha \cosh \beta) \} , 
\ee
\be \label {e15}
e^{-\Phi} = {\Delta^{1/2} \over r^{D-4}} , 
\ee
\bea \label{e16}
A^{\ab}_{ t} &=& -{n^{\ab} \over 2 \sqrt 2 \Delta} C \sinh \alpha
\{ r^{D-4} \cosh\beta + {1\over 2} C (\cosh\alpha + \cosh \beta) \}
\nonumber \\
&& \qquad \qquad \hbox{for} \qquad 1 \le a \le (10-D) \, , \nonumber \\
&=& -{p^{(a-10+D)} \over 2 \sqrt 2 \Delta} C \sinh \beta
\{ r^{D-4} \cosh\alpha + {1\over 2} C (\cosh\alpha + \cosh \beta) \}
\nonumber \\
&& \qquad \qquad \hbox{for} \qquad (10-D)+1 \le a \le (20 - 2D) \, , 
\nonumber \\
\eea
\bea \label{e17}
A^{\ab}_{ D-1} &=& - {n^{\ab} \over 2\sqrt 2 \Delta} C \sinh \alpha 
\{r^{D-4}
+ {1\over 2} C \cosh \beta (\cosh\alpha + \cosh \beta) \} \nonumber \\
&& \qquad \qquad \hbox{for} \qquad
1\le a \le (10-D) \, , \nonumber \\
&=&  {p^{(a-10+D)} \over 2\sqrt 2 \Delta} C \sinh \beta \{ r^{D-4}
+ {1\over 2} C \cosh \alpha (\cosh\alpha + \cosh \beta) \} \nonumber \\
&& \qquad  \qquad \hbox{for} \qquad
(10-D)+1 \le a \le (20-2D) \, , \nonumber \\
\eea
\be \label{e18}
M_D = I_{20-2D} + \pmatrix{ P nn^T & Q np^T \cr Q pn^T & P pp^T \cr} \, ,
\ee
where,
\be \label{e19}
\Delta = r^{2(D-4)} + Cr^{D-4} ( 1 + \cosh\alpha \cosh\beta) + {C^2 \over 4}
(\cosh\alpha + \cosh\beta)^2 \, ,
\ee
\be \label{e19a}
P = {C^2 \over 2\Delta} \sinh^2 \alpha \sinh^2 \beta \, ,
\ee
\be \label{e19b}
Q = - C \Delta^{-1} \sinh\alpha \sinh\beta \{ r^{D-4} + {1\over 2} C
(1 + \cosh\alpha \cosh \beta) \} \, .
\ee
with $n^{(a)}$, $p^{(a)}$ being the components of $(10-D)$-dimensional 
unit vectors. 
$A_{\mu}$'s in eqns. (\ref{e16}), (\ref{e17})
are the gauge fields appearing due to the Kaluza-Klein (KK)
reductions of the ten dimensional metric and the 2-form antisymmetric
tensor coming from the NS-NS sector, $B_{\mu \nu}$ is 
the NS-NS 2-form field and $\Phi$ is the dilaton in the D-dimensional
space-time. In the above configuration, $C$ is 
a constant related to tension of the string. The matrix $M_D$ parameterizes
the moduli fields. The exact form of this parameterization 
depends on the form of the $O(10-D, 10-D)$ metric used. 
The above solution has been written for a diagonal
metric of the form: 
\be
 L_D = \pmatrix{ -I_{10-D} & \cr & I_{10-D}}. \label{eld}
\ee
One sometimes also uses an off-diagonal metric convention
(as in eqn.(\ref{decomp}) below:
\be
L = \pmatrix{ & I_{10-D} \cr I_{10-D} & }. \label{el}
\ee
These two conventions are however related by:
\be
 L_D = \hat{P} L \hat{P}^T, \>\>\> M_D = \hat{P} M \hat{P}^T, \label{eleld}
\ee
where 
\be
   \hat{P} = {1\over \sqrt{2}}\pmatrix{-I_{10-D} & I_{10-D} \cr 
I_{10-D} & I_{10-D}}. \label{hatp}
\ee 

The gauge fields in two conventions are related as:
\be
  \pmatrix{A^1_{\mu} \cr A^2_{\mu}} = 
\hat{P} \pmatrix{\hat{A}^1_{\mu} \cr \hat{A}^2_{\mu}}, 
            \label{aredef}
\ee
with $A^{1,2}_{\mu}$'s in the above equation
being $(10-D)$-dimensional columns consisting 
of the gauge fields $A_{\mu}$'s defined in (\ref{e16}-\ref{e17}), 
and coming from the left and the right-moving sectors of string theory.  

\subsection{Decompactified Solutions for $\beta=0$, $\alpha \neq 0$ 
and $\alpha = - \beta$}

We now discuss the decompactification of the 
D-dimensional solutions (\ref{e13})-(\ref{e19b}) to $D=10$. 
As stated earlier, we now restrict ourselves to specific values of
the parameters: $\alpha = -\beta$ and $\beta =0, \alpha \neq 0$ for
algebraic simplifications. These special cases, in particular the later
possibility, encompasses all the nontrivialities of our construction. 
We also restrict ourselves to $D=9$ for constructing $D0-D2$ bound state. 
First we start with the seed solution (with $\beta =0$,
$\alpha$ arbitrary) given as:
\bea
ds^{2} &=& {1\over \cosh^2{\alpha\over 2} e^{-E} -\sinh^2{\alpha\over
2}} (-dt^2 +(dx^{D-1})^2) \cr
& \cr
&+& {\sinh^2{\alpha\over 2} (e^{-E}-1)\over
(\cosh^2{\alpha\over 2} e^{-E} -\sinh^2{\alpha\over
2})^2} (dt+dx^{D-1})^2 + \sum_{i=1}^{D-2} dx^i dx^i , \cr
& \cr
B_{(D-1)t} &=& {\cosh^2{\alpha\over 2}(e^{-E}-1)\over \cosh^2{\alpha\over
2} e^{-E} -\sinh^2{\alpha\over2}} , \cr
& \cr
A^{(1)}_{D-1} &=& A^{(1)}_t = - {1\over {2\sqrt{2}}}\times
{\sinh\alpha (e^{-E}-1)\over
\cosh^2{\alpha\over 2} e^{-E} -\sinh^2{\alpha\over
2}} , \cr 
& \cr
\Phi &=& -\ln(\cosh^2{\alpha\over 2}e^{-E} -\sinh^2{\alpha\over 2}) , 
\label{betaeq0}
\eea
with $e^{-E}$ being the Green function in the 5-dimensional 
transverse space:
\be
e^{-E} = (1 + {C\over r^5}) ,  \label{green}
\ee
and constant $C$ determines the string tension. 

Now, to construct delocalized solutions in $D=10$, we decompactify the 
above solution back to ten dimensions. 
The decompactification exercise is done following a set of notations 
given in \cite{senijmp}. When restricted to the NS-NS sector of 
type II theories, they can be written as:
\bea\label{decomp}
&& \hat {G}_{a b}  = G^{\ten}_{[a+(D-1), b+(D-1)]}, 
\quad  \hat B_{a b}  =
B^{\ten}_{[a+(D-1), b+(D-1)]}, 
\nonumber \\
&& \hat{A}^{(a)}_{\bar{\mu}}  = {1\over 2}\hat G^{ab} 
G^{\ten}_{[b+(D-1),\bar{\mu}]}, \nonumber \\
&&  \hat{A}^{(a+(10-D))}_{\bar{\mu}} = {1\over 2}
B^{\ten}_{[a+(D-1), \bar{\mu}]} - \hat B_{ab} A^{(b)}_{\bar{\mu}}, 
\nonumber \\
&& G_{\bar{\mu}\bar{\nu}} = G^{\ten}_{\bar{\mu}\bar{\nu}} 
- G^{\ten}_{[(a+(D-1)), \bar{\mu}]} 
G^{\ten}_{[(b+(D-1)), \bar{\nu}]} \hat
G^{ab}, \nonumber \\
&& B_{\bar{\mu}\bar{\nu}} = 
B^{\ten}_{\bar{\mu}\bar{\nu}} - 4\hat B_{ab} A^{(a)}_{\bar{\mu}}
A^{(b)}_{\bar{\nu}} - 
2 (A^{(a)}_{\bar{\mu}} A^{(a+(10-D))}_{\bar{\nu}} - A^{(a)}_{\bar{\nu}} 
A^{(a+(10-D))}_{\bar{\mu}}),
\nonumber \\
&& \Phi = \Phi^{\ten} - {1\over 2} \ln\det \hat G, \quad 
\quad 1\le a, b \le 10-D, \quad
0\le {\bar{\mu}}, \bar{\nu} \le (D-1).
\eea

We now start with the nine-dimensional ($D=9$) solution in 
(\ref{betaeq0}) and 
following the Kaluza-Klein (KK) compactification mechanism summarized above,
write down the solution directly in ten dimensions\cite{kumar}. 
For $\beta = 0$, only nonzero background fields are then given by
\bea \label{10dg}
ds^2 &=&  {1\over {\cosh^2 {\alpha\over 2} e^{-E} - \sinh^2 {\alpha\over 2}}}
         ( -dt^2 + (dx^8)^2 ) \cr
      &+& {\sinh^2{\alpha\over 2}(e^{-E} - 1)\over 
        {\cosh^2 {\alpha\over 2}e^{-E} - \sinh^2 {\alpha\over 2}}}
        (dt + dx^{8})^2  \cr 
& \cr
&+& {{\sinh \alpha (e^{-E} - 1)}\over 
                  {\cosh^2 {\alpha\over 2}e^{-E} - \sinh^2 {\alpha\over 2}}}
     dx^9 (dt + dx^8) +  \sum_{i=1}^{7} dx^i dx^i + (dx^9)^2,
\eea
\bea \label{10db}
B_{8 t}  &=& {{\cosh^2 {\alpha\over 2} (e^{-E} - 1)}
              \over {\cosh^2 {\alpha\over 2}e^{-E} - 
\sinh^2 {\alpha\over 2}}},\cr
B_{9 t} &=& - {\sinh\alpha\over 2}{{(e^{-E} - 1)}\over 
{\cosh^2 {\alpha\over 2}e^{-E} - \sinh^2 {\alpha\over 2}}} = B_{98}.
\eea
The dilaton in ten dimensions remains same as the one in 
(\ref{betaeq0}):
\be \label{10phi} 
\Phi^{(10)} = - \ln ({\cosh^2 {\alpha\over 2} e^{-E}
- \sinh^2 {\alpha\over 2}}).
\ee

For $\alpha= -\beta$, on the other hand, we have $D=9$ solutions given
by a metric:
\be
ds^2 = -{1\over {1 + {{C \cosh^2\alpha}\over r^5}}} {dt}^2 +
        {1\over {1 + {C\over r^5}}} {(dx^8)}^2 + \sum_{i=1}^{7}
        dx^i dx^i.  \label{al=beg}
\ee
The only non-zero component of the antisymmetric tensor is of the
form
\be
B_{t 8} = - {C \cosh\alpha\over 2}\left[{1\over {(r^5 + C)}} 
+ {1\over {(r^5 + C \cosh^2\alpha)}}\right].
          \label{al=beb}
\ee
We also have a nontrivial modulus parameterizing the $O(1, 1)$
matrix $M_D$ in eqn.(\ref{e18}):
\be
   \hat{G}_{99} \equiv \hat{g} = {{1+ {C \cosh^2\alpha\over r^5}}\over
              {1 + {C\over r^5}}}. 
           \label{hatg}
\ee
The two gauge fields appearing in equations (\ref{decomp})  
for $D=9$ are of the form:
\be
\hat{A}^1_t = {{C \sinh\alpha \cosh\alpha}\over { 2 
(r^5 + C \cosh^2\alpha)}}, \>\>\> \hat{A}^1_8 = 0, 
\ee
\be
\hat{A}^2_t = 0, \>\>\>
\hat{A}^2_8 = {-{C \sinh\alpha}\over { 2 
(r^5 + C)}}. 
\ee

The decompactified solution for $\alpha = -\beta$ case in $D=10$ 
is given by:

\begin{eqnarray}
ds^2 &=& - {(1 - {C \over r^5}{\rm sinh}^2 {\alpha})
     \over {1 + {C \over r^5}}}(dt)^2   
    + {1 + {C \over r^5}{\rm cosh}^2{\alpha}\over {1 + {C \over
      r^5}}}({dx^9})^2 \cr
& \cr  
&+& {{2 C \over r^5}{\rm cosh}{\alpha}{\rm sinh}{\alpha}
       \over{1 + {C \over r^5}}}{dx^9}dt + {1 \over{1 + {C \over
         r^5}}}({dx^8})^2 + \sum_{i=1}^7 (dx^i)^2,
\label{fstring}
\end{eqnarray}
antisymmetric NS-NS $B_{\mu \nu}$:
\begin{equation}
B_{98} = - C {{{\rm sinh}\alpha}\over{r^5 + C}}, \>\>\>
B_{t8} = - C {{{\rm cosh}\alpha}\over{r^5 + C}}, 
\label{b_munu}
\end{equation} 
and by the dilaton:
\begin{equation}
\Phi^{(10)}= - {\rm ln} (1 + {C \over r^5}). 
\label{iiphi}
\end{equation} 
Before applying $SL(2, Z)$ transformation to the solutions in eqns.
(\ref{10dg})-(\ref{10phi}) and (\ref{fstring})-(\ref{iiphi}),
we now describe the general construction of 
$Dp - D(p+2)$ bound states starting from $D$- strings of type IIB\cite{myers}.
  
\sect{\bf Construction of $Dp-D(p+2)$ bound states}

In this section we first review the construction of $D0 - D2$ 
non-threshold bound states in type IIA string theories from 
the (neutral)  $D$-strings in type IIB. Our starting point
in this case is the $D$-string solution, smeared (delocalized) along
a transverse direction $x \equiv x^9$. This solution is given as:
\begin{eqnarray}
ds^2 &=& \sqrt{H} \left( {{-dt^2 + dy^2}\over H} +  dx^2
+ \sum_{i=1}^7 (dx^i)^2 \right), \cr
& \cr
A^{(2)} &=& \pm \left( {{1\over H} - 1 }\right) dt \wedge dy, \cr
& \cr
e^{\Phi_b^{(10)}} &=& H.
\label{d1neutral}
\end{eqnarray}
\begin{equation}
H = 1 + {C\over r^5}.
\label{eh}
\end{equation}
is the solution of the Greens' function equation:
\begin{equation}
\Box H = C A_{N-1} \prod_{i=1}^N \delta(x^i),
\label{greenfunction}
\end{equation}
where $N$ indicates the transverse directions which are not smeared
and $A_N$ is the
area of $S^N$ orthogonal to the brane. 
To compare with the solutions in section-2, one identifies $H = e^{-E}$.  
Then a rotation is performed between $x$ and $y$ directions:
\begin{eqnarray}
\pmatrix{x\cr y} = \pmatrix{ \cos\phi & -\sin\phi \cr
                         \sin\phi & \cos\phi} 
                   \pmatrix{ \tilde{x}\cr \tilde{y}}
\label{rotation}
\end{eqnarray}
to mix the longitudinal and transverse coordinates of the above
solution. 
The solution in eqn.(\ref{d1neutral}) then transforms to:
\begin{eqnarray}
ds^2 &=& \sqrt{H}\Big[ {-dt^2\over H} + ({\cos^2\phi\over H} 
+ \sin^2\phi) d\tilde{y}^2 +({\sin^2\phi\over H} 
+ \cos^2\phi) d\tilde{x}^2 \cr
& \cr
&+& 2 \cos\phi \sin\phi ({1\over H} - 1)d\tilde{y}d\tilde{x} 
+ \sum_{i=2}^8 (d {x}^i)^2 \Big],  \cr
& \cr
A^{(2)} &=& \pm \left( {{1\over H} - 1 }\right) dt 
\wedge (\cos\phi d\tilde{y} + \sin\phi{d\tilde{x}} ), \cr
& \cr
e^{\Phi_b^{(10)}} &=& H.
\label{d1rot}
\end{eqnarray}
Finally, a  T-duality transformation \cite{hull,myers} on 
coordinate $\tilde{x}$ gives 
the following classical solution in the type IIA theory:
\begin{eqnarray}
ds^2 &=& \sqrt{H} \left( {-{dt^2\over H}}+
{{d{\tilde x}^2 + d{\tilde y}^2}\over {1+ (H-1)\cos\phi}} +  
\sum_{i=2}^8 (dx^i)^2 \right), \cr
& \cr
A^{(1)} &=& \pm \left( {1\over H} - 1 \right)\sin\phi dt, \cr
& \cr
A^{(3)} &=& \pm { (H-1)\cos \phi \over{1 + (H-1)\cos^2 \phi}}
dt\wedge d\tilde{x} \wedge d\tilde{y}, \cr
& \cr
B^{(a)} &=& { (H-1)\cos\phi \sin\phi \over{1 + (H-1)\cos^2 \phi}}
d\tilde{x} \wedge d\tilde{y}, \cr
& \cr
e^{\Phi_a^{(10)}} &=& { H^{3\over 2} \over{1 + (H-1)\cos^2 \phi}}.
\label{d0-d2-2a}
\end{eqnarray}
Solution in eqn. (\ref{d0-d2-2a}) can be interpreted as a $D0 - D2$ bound
state. The $D0$ and $D2$ charge densities carried by the above
bound state solution are given as:
\begin{eqnarray}
Q_0 =  5 C \sin\phi A_6, \cr
Q_2 =  5 C \cos\phi A_6, \cr
\label{02charge}
\end{eqnarray}
where $A_6$ is the  area of $S^6$ orthogonal to the brane.
The ADM mass \cite{lu,myers} is defined by the expression:
\begin{equation}
m = \int \sum_{i=1}^{9-p} n^i
\left[\sum_{j=1}^{9-p} (\partial_j h_{i j} - \partial_i h_{j j}  )
 - \sum_{a=1}^p \partial_i h_{aa} \right] r^{8 - p} d\Omega,
\label{adm-mass}
\end{equation}
where $n^i$ is a radial unit vector in the transverse space and 
$h_{\mu \nu}$ is the deformation of the Einstein-frame metric
from flat space in the asymptotic region. 
We also should mention here that in order to write mass and charges, we
have tuned the gravitational constant to a suitable value.
The ADM mass density in the present context is found to be:
\begin{equation}
m_{0, 2} = 5 C A_6.
\label{02mass}
\end{equation}
Mass and charge densities given in eqns. (\ref{02charge}) and (\ref{02mass})  
above satisfy the BPS condition:
\begin{equation}
(m_{0,2})^2 = (Q_0^2 + Q_2^2).
\label{mass-charge}
\end{equation}

To generalize the results to other D-brane bound states, we 
smear one more transverse direction, $x^i \sim x^7$ in eqn. 
(\ref{d0-d2-2a}). Finally applying $T$-duality along 
this direction, one is able to construct a $D1-D3$ bound state. 
In the dualized (IIB) theory, one can write down  
all the field  components trivially. One finds an exact matching 
with the $D1-D3$ solution of \cite{myers}. In particular, 
matching of the $4$-form field can be seen by  using the 
identity: 
\begin{eqnarray}
{A^{(4)}} &\equiv & {(H - 1) \cos{\phi}\over {1 + (H - 1)\cos^2{\phi}}} 
- {1 \over2}{(H - 1)\over H} {\sin{\phi}(H - 1) \cos{\phi}\sin{\phi}
\over{1 + (H - 1)\cos^2{\phi}}}\cr
& \cr
&=& {(H - 1) \cos{\phi}\over 2 H}
\left[{1 + {H \over{1 + (H - 1)\cos^2{\phi}}}}\right].
\label{d1-d3-myers}
\end{eqnarray} 
We give the final $D1 - D3$ solution \cite{myers}
for completeness as well as for
later use:
\begin{eqnarray}
ds^2 &=& \sqrt{H}\Big\{{- d t^2 +(d y^2)^2 \over H} + 
{ d \tilde y^2 + d \tilde x^2 \over 1 + (H-1) \cos^2 \phi} + d r^2\cr
& \cr
&+& r^2(d \theta^2 + \sin^2 \theta(d \phi_1^2 +
\sin^2 \phi_1 (d \phi_2^2 + \sin^2 \phi_2( d \phi_3^2 
+\sin^2 \phi_3 d \phi_4^2))))
\vphantom{1\over H}\Big\}\cr
& \cr
A^{(4)} &=& \mp{\cos \phi\over2}{H-1 \over H}
\left(1+  {H \over
1 + (H-1) \cos^2 \phi}\right)\times d t \wedge d \tilde y \wedge
d y^2 \wedge d \tilde x \cr
& \cr
&\pm& 4 C \cos\phi\,\sin^4\theta\sin^3\phi_1\sin^2\phi_2
\cos\phi_3 d\phi\wedge d\phi_1\wedge d\phi_2\wedge d\phi_4, \cr
& \cr
A^{(2)} &=&\pm {H-1 \over H}\, \sin \phi\, d t \wedge d y^2,\cr
& \cr
B^{(b)} &=& { (H -1) \cos \phi \sin \phi 
\over  1 + (H-1) \cos^2 \phi} d \tilde x \wedge d \tilde y, \cr
& \cr
e^{\Phi_b^{(10)}} &=&\, {H \over  1 + (H-1) \cos^2 \phi},
\label{d1-d3-myers2}
\end{eqnarray}
where $H = 1 + {{C \over r^4}}$.

One can now repeat this process to generate all the bound states
of $Dp-D(p+2)$ type in a similar manner. 
We now apply the general procedure described above 
to the charged macroscopic string solutions of section-2.

\sect{\bf Generalized $(Dp - D(p+2))$ bound state}
In this section we construct   
non-threshold bound states which are the 
generalization of the $D0-D2$ bound state presented in the previous
section.  Here we use the  
$\alpha =  -\beta$ solutions of section-2 and postpone the 
discussion of $\beta = 0, \alpha \neq 0$ solutions to  
section-5.  

\subsection{Generalization of D0-D2 Bound States}

 The delocalized elementary
string solution is given in eqns.(\ref{fstring})-(\ref{iiphi}).
A delocalized $D$-string in $D=10$ can be generated from this
solution by an application of $S$-duality transformation \cite{schwarz}
which transforms an elementary string into a $D$-string. 
The metric, antisymmetric 2-form
$(B_{\mu \nu})$ and the dilaton for the delocalized 
$D$-string solution are given by:
\begin{eqnarray}
ds^2 &=& - {{1 - {C\over r^5}{\rm sinh^2}{\alpha}}\over{\sqrt{1 +
      {C\over r^5}}}}{(dt)}^2 + 
{{1 + {C\over r^5}{\rm cosh^2}{\alpha}}\over{\sqrt{1 +{C\over
        r^5}}}}{(dx^9)}^2 \cr
& \cr
&+& {2 {C\over r^5} {\rm sinh}{\alpha}{\rm cosh}{\alpha}\over{\sqrt{ 1
      + {C\over r^5}}}} {dx^9}{dt} + {1 \over{\sqrt{1 + {C\over
    r^5}}}}{(dx^8)}^2 \cr
&\cr
&+& {\sqrt{ 1 + {C\over r^5}}}\sum_{i=1}^7
{(dx^i)}^2,
\label{iibm}
\end{eqnarray}
\begin{equation}
B^{(2)}_{9 8} = B_{9 8},~B^{(2)}_{t 8} = B_{t 8}, 
~e^{{\Phi}_b^{(10)}} = {{ 1 + {C\over r^5}}},
\end{equation}
with $B_{9 8}$ and $B_{t 8}$ as given in eqn. (\ref{b_munu}) and 
the superscript on $B$ denotes the R - R nature of the 2-form
antisymmetric tensors. The next step of our construction is to apply
rotation in $(x^9- x^8)$- plane by an angle $\phi$ which gives the following
configuration:
\begin{eqnarray}
ds^2 &=& -{{1 - {C\over r^5}{\rm sinh^2}{\alpha}\over{\sqrt{1 + {C\over
          r^5}}}}{dt^2} + {1 + {C\over r^5}{\rm cosh^2}{\alpha}{\rm
      cos^2}{\phi}\over{\sqrt{1 + {C\over r^5}}}}}{(d \tilde x^9)}^2 
\cr
& \cr
&+& {1 + {C\over r^5}{\rm cosh^2}{\alpha}{\rm
      sin^2}{\phi}\over{\sqrt{1 + {C\over r^5}}}}{(d \tilde x^8)}^2 
+ {2 {C\over r ^5}{\rm cosh}{\alpha}{\rm sinh}{\alpha}{\rm
    cos}{\phi}\over{\sqrt{1 + {C\over r^5}}}}{d \tilde x^9}{dt}\cr
& \cr
&-& {2 {C\over r ^5}{\rm cosh}{\alpha}{\rm sinh}{\alpha}{\rm
    sin}{\phi}\over{\sqrt{1 + {C\over r^5}}}}{d \tilde x^8}{dt} -
{{2{C\over r^5}{\rm cosh^2}{\alpha}{\rm sin}{\phi}{\rm
    cos}{\phi}}\over{\sqrt{1 +{C\over r^5}}}}{(d \tilde x^9)}{d
\tilde x^8}\cr
& \cr
&+& {\sqrt{1 +{C\over r^5}}}\sum_{i=1}^7 {(dx^i)}^2,
\end{eqnarray}
\begin{eqnarray}
B^{(2)}_{\tilde 8 t} &=& {C \over {C + r^5}}{\rm cosh}{\alpha}{\rm
  cos}{\phi},\cr
& \cr
B^{(2)}_{\tilde 9 t} &=& {C \over {C + r^5}}{\rm cosh}{\alpha}{\rm
  sin}{\phi},\cr
& \cr
B^{(2)}_{\tilde 9 \tilde 8} &=& {-C \over {C + r^5}}{\rm
sinh}{\alpha}.
\end{eqnarray}
Finally we apply T-duality along 
the $\tilde x^9$-direction. By following the prescription as given in
\cite{myers, hull}, we end up with the structure for the
metric, NS- NS $B_{\mu \nu}$, as well as 1- form and 3- form fields 
of type IIA theory:
\begin{eqnarray}
d{\cal{S}}^2 &=& -{1 + {C\over r^5}(1 - {\rm cosh^2}{\alpha}\sin^2{\phi}) 
\over{\sqrt{1 + {C\over r^5}}(1 + {C \over r^5}\cosh^2{\alpha}
\cos^2{\phi})}}{dt^2} \cr
& \cr
&+&{{\sqrt{1 + {C\over r^5}}}\over{1 + {C\over r^5}}
 {\rm cosh^2}{\alpha}{\rm cos^2}{\phi}}{(d \tilde x^9)}^2 
+{{1 + {C\over r^5}{\rm cosh^2}{\alpha}}\over{\sqrt{1 +{C\over
      r^5}}}(1 + {C\over r^5}{\rm cosh^2}{\alpha}{\rm cos^2}
{\phi})}{(d \tilde x^8)}^2 \cr
& \cr
&-& {{C \over r^5}{\rm sinh}{\alpha}{\rm cosh}{\alpha}{\rm
  sin}{\phi}\over{\sqrt{{1 + {C\over r^5}}}
(1 + {C\over r^5}{\rm cosh^2}{\alpha}
{\rm cos^2}{\phi})}} dt{d \tilde x^8}
+ {\sqrt{1 + {C \over r^5}}}\sum_{i=1}^7 {(dx^i)}^2,
\label{d0-d2-metric}
\end{eqnarray}
\begin{eqnarray}
e^{ \Phi_b^{(10)}} &=& {{(1 +{C\over r^5})^{3/2}}\over{1 + {C\over
r^5}{\rm
    cosh^2}{\alpha}{\rm cos^2}{\phi}}},\cr
& \cr
{\cal{A}}_t &=& {C \over{C + r^5}}{\rm cosh}{\alpha}{\rm
  sin}{\phi},~~~~{\cal{A}}_{\tilde x^8} = {-C\over {C + r^5}}
{\rm sinh}{\alpha}\cr
& \cr
{\cal{B}}_{\tilde 9 t} &=& {{{-C \over r^5}{\rm sinh}{\alpha}{\rm
  cosh}{\alpha}{\rm cos}{\phi}}\over{1 + {C \over r^5}{\rm
    cosh^2}{\alpha}{\rm cos^2}{\phi}}},~~~~ 
{\cal{B}}_{\tilde 9 \tilde 8}  =  {{{C \over r^5}{\rm sin}{\phi}{\rm
  cos}{\phi}{\rm cosh^2}{\alpha}}\over{1 + {C \over r^5}{\rm
    cosh^2}{\alpha}{\rm cos^2}{\phi}}},\cr
& \cr 
{\cal{A}}_{\tilde 9 t \tilde 8} &=& - {C \over r^5}{\rm cosh}{\alpha}{\rm
  cos}{\phi}\over{1 + {C \over r^5}{\rm cosh^2}{\alpha}
 {\rm cos^2}{\phi}}.
\label{iiamat}
\end{eqnarray}
This solution is a generalization of the 
$D0-D2$ bound state, where in addition we have turned on 
NS-NS $2$-form as well. To show that this indeed represents a
non-threshold $1/2$ BPS state, in the next sub-section, we 
explicitly verify the mass-charge relation of these bound states.
\subsection{Mass-Charge Relationship}

To show the BPS nature of our solution, we perform a dimensional 
reduction of our solution along $\tilde{x}^8$ and 
$\tilde{x}^9$ directions. By dimensional reduction, one avoids any 
ambiguity that may arise due to the presence of 
purely spatial components of the $p$-form fields in 
equation (\ref{d0-d2-metric}), (\ref{iiamat}) above. 
As a result, all the nonzero charges
in our case arise from the temporal part of $1$-form field components 
only, in this eight dimensional theory \cite{roy}. They are
$A^1_t \sim {\cal{A}}_t$, $A^2_t \sim {\cal{B}}_{\tilde 9 t}$, 
$A^3_t \sim {\cal{A}}_{\tilde 9 t
  \tilde 8}$ from the components of the $p$-form 
fields, as well as off-diagonal component 
${g_{t \tilde 8}}/{g_{\tilde 8\tilde 8}}$ of the metric.
The charges associated with these field strengths and metric
components can be read from the solutions 
(\ref{d0-d2-metric}), (\ref{iiamat}). They are :
\begin{eqnarray}
Q_1 &=& 5 C {\rm cosh}{\alpha}~{\sin}{\phi},\cr
& \cr
Q_2 &=& - 5 C {\rm sinh}{\alpha}~{\cosh}{\alpha}~{\rm cos}{\phi}, \cr
& \cr
Q_3 &=&  - 5 C {\rm cosh}{\alpha}~\cos \phi, \cr
& \cr
P &=& 5 C {\rm sinh}{\alpha}~{\cosh}{\alpha}~{\rm sin}{\phi},
\label{02charges}
\end{eqnarray} 
where $Q_i$'s are the charges corresponding to 
the field strengths of $A^i_t$ that we just defined 
and $P$ can be interpreted as the momentum along
$\tilde{x}^8$ direction in the ten-dimensional theory.
The mass-density of our bound state can be computed
using (\ref{adm-mass}) and is given by
\begin{equation}
m_{(0, 2)} = 5 C {\rm cosh^2}{\alpha}.
\label{mtwo}
\end{equation} 
Comparing (\ref{02charges}) and (\ref{mtwo}), we get the standard 
BPS condition as:
\begin{equation}
(m_{0, 2})^2 = {Q^2}_1 + {Q^2}_2 + {Q^2}_3 + P^2.
\end{equation}
This, in turn, implies the supersymmetric nature of the bound state.

\subsection{Further Generalization to $Dp-D(p+2)$}

We now obtain a generalization of the $D1-D3$ bound state solution 
presented in eqn. (\ref{d1-d3-myers2}) by applying T-duality along 
$x^7$ direction on the generalized $D0-D2$ solution
(\ref{d0-d2-metric}), (\ref{iiamat}) presented earlier in this section. 
The final result is given by:
\noindent
\begin{eqnarray}
d{\cal{S}}^2 &=& -{1 + {C\over r^4}(1 - \cosh^2{\alpha}\sin^2{\phi}) 
\over{\sqrt{1 + {C\over r^4}}(1 + {C\over r^4}\cosh^2{\alpha}
\cos^2{\phi})}}{dt^2} + {{\sqrt{1 + {C\over r^4}}}\over{1 + {C\over r^4}}
 {\rm cosh^2}{\alpha}{\rm cos^2}{\phi}}{(d \tilde x^9)}^2 \cr
& \cr
&+&{{1 + {C\over r^4}{\rm cosh^2}{\alpha}}\over{\sqrt{1 +{C\over
      r^4}}}(1 + {C\over r^4}{\rm cosh^2}{\alpha}{\rm cos^2}
{\phi})}{(d \tilde x^8)}^2 + {1\over\sqrt{1 + {C\over
    r^4}}}{(dx^7)}^2\cr
& \cr 
&-& {{C \over r^4}{\rm sinh}
{\alpha}{\rm cosh}{\alpha}{\rm sin}{\phi}\over{\sqrt{{1 + {C\over r^4}}}
(1 + {C\over r^4}{\rm cosh^2}{\alpha} {\rm cos^2}{\phi})}} 
dt{d \tilde x^8} + {\sqrt{1 + {C \over r^4}}}\sum_{i=1}^6 {(dx^i)}^2,\cr
& \cr
{\cal{A}}^{(2)}_{7 t} &=& {C\sin \phi \cosh \alpha\over(r^4 +
C)},~~~~~ {\cal{A}}^{(2)}_{7 \tilde 8} =
- {C\sinh\alpha\over(r^4 + C)}, \cr
& \cr
{\cal{A}}^{(4)}_{\tilde 9 t \tilde 8 7 } 
&=& - {({C\over r^4} \cosh \alpha \cos \phi) \over { 2( 1 + {C\over r^4})}}
\left[1 + {{1 + {C\over r^4} \over{{1 + {C\over r^4}}
\cosh^2\alpha\cos^2\phi}}}\right],\cr
& \cr
{\cal{B}}_{\tilde 9 t} &=& {{{-C \over r^4}{\rm sinh}{\alpha}{\rm
cosh}{\alpha}{\rm cos}{\phi}}\over{1 + {C \over r^4}{\rm
cosh^2}{\alpha}{\rm cos^2}{\phi}}},~~~~ 
{\cal{B}}_{\tilde 9 \tilde 8}  =  {{{C \over r^4}{\rm sin}{\phi}{\rm
cos}{\phi}{\rm cosh^2}{\alpha}}\over{1 + {C \over r^4}{\rm
cosh^2}{\alpha}{\rm cos^2}{\phi}}},\cr
& \cr
e^{{\Phi}_{b}^{(10)}} &=& {1 + {C\over r^4}\over{1 + {C\over
r^4}\cosh^2\alpha\cos^2\phi}}.
\end{eqnarray}
Once again, for the special case $\alpha = 0$, our solution
reduces to the one in \cite{myers}. We have therefore presented 
a generalization of the $D1-D3$ bound state to include new 
charges and momenta. The BPS nature of the new solution can be 
again examined by looking at the leading behavior of the 
gauge fields when the above solution is reduced along all its
isometry directions: $x^7, \tilde{x}^8, \tilde{x}^9$. They are:
\begin{eqnarray}
Q_1 &=& 4 C \cosh \alpha \sin \phi, \cr
& \cr 
Q_2 &=& - 4 C \cosh \alpha \cos \phi, \cr
& \cr
Q_3 &=& - 4 C \sinh \alpha \cosh \alpha \cos \phi, \cr
& \cr
P &=& - 4 C \sinh \alpha \cosh \alpha \sin \phi,
\end{eqnarray}
where $Q_1$, $Q_2$, $Q_3$ and $P$ are the charge 
associated with  ${\cal{A}}^{(2)}_{7 t}$, ${\cal{A}}^{(4)}_{\tilde 9 t 
\tilde 8 7 }$, ${\cal{B}}_{\tilde 9 t}$ and $G_{t \tilde 8}$, respectively.

In order to compute the ADM mass-density, we find:
\begin{eqnarray}
h_{77} &=& {C\over r^4}\left({-3\over4} + {1\over 4}\cosh^2 \alpha \cos^2
  \phi\right),~h_{\tilde 8 \tilde 8} = {C\over
  r^4}\left(\cosh^2\alpha - {3\over 4} \cosh^2 \alpha \cos^2 \phi
  -{3\over 4}\right),\cr
& \cr
h_{\tilde 9 \tilde 9} &=& {C \over r^4}\left({1\over 4} - {3 \over
    4}\cosh^2\alpha\cos^2 \phi\right),~h_{i j} = {C\over
  r^4}\left({1 \over 4} + {1\over 4}\cosh^2 \alpha \cos^2
  \phi\right){\delta}_{i j},
\end{eqnarray} 
with $h_{i j}$'s being the deformations of the Einstein metric
above flat background. One then gets, using (\ref{adm-mass}), 
the mass-density of the $D1-D3$ system as:
\begin{eqnarray}
m_{1,3} = 4 C \cosh^2 \alpha.
\end{eqnarray}
We therefore once again have:
\begin{equation}
m_{1,3}^2 = Q_{1}^2 + Q_{2}^2 + Q_{3}^2 + P^2,
\end{equation}
showing the BPS nature of the new bound state. Further generalization 
to higher $Dp-D(p+2)$ bound states can similarly be worked out.
We therefore skip the details. 

\sect{\bf $\beta = 0, \alpha \neq 0$ Solutions and Generalization 
to $(p, q)$- Strings}

In this section, we discuss even more nontrivial examples, using 
$\beta =0, \alpha \neq 0$ cases, discussed in 
section-2. We now also perform further generalization by applying 
$T$-duality on delocalized $(p, q)$-strings rather than 
$(0, 1)$ or D-strings. 

\subsection{Construction of SL(2, Z) Multiplets}
The delocalized elementary string solution in ten-dimensions
for $\beta = 0, \alpha \neq 0$ situation  is given 
in eqns.(\ref{10dg})-(\ref{10phi}). This solution represents a string
along $x^8$, which is delocalized along $x^9$. One can use it to 
construct generalized bound states in the manner described in the 
last section. We, however, make further generalization by 
using the $SL(2,Z)$ symmetry of type IIB string theories in $D=10$. 
We can easily construct 
a $(p,q)$ multiplet of delocalized string from (\ref{10dg})-(\ref{10phi}). 
The procedure of constructing such configuration is discussed 
in \cite{schwarz}. Instead of giving the detail, we write down the 
final form of the configuration.

The metric is given by:
\begin{eqnarray}
ds^2 &=& {A\over \Delta}\Big(- [1 - {\rm sinh}^2 {\alpha\over 2} (e^{-E}
-1)]
dt^2 + [1 + {\rm sinh}^2 {\alpha\over 2} (e^{-E} -1)] (dx^8)^2\nonumber\\
&+& \Delta (dx^9)^2 
+  {\rm sinh}\alpha (e^{-E} -1)dx^9 dt +
{2} {\rm sinh}^2{\alpha\over 2} (e^{-E} -1) dt dx^8
\nonumber\\
&+&  {\rm sinh}\alpha (e^{-E} -1) dx^8 dx^9 +
\Delta \sum_{i=1}^7 dx^i dx^i \Big),
\label{pqdeloca}
\end{eqnarray}
where we have defined 
\begin{equation} 
\Delta = {\rm cosh}^2{\alpha\over 2}e^{-E} - {\rm sinh}^2{\alpha\over 2},
~~{\rm and}~~ A = {{\sqrt{p^2 + q^2 \Delta}}\over{{\sqrt{p^2 + q^2}}}}.
\end{equation}
Furthermore, the NS-NS and R-R two forms are given by
\begin{eqnarray}
B^{(1)}_{8t} = {p\over{\sqrt{p^2 + q^2}}} B_{8t},
~~B^{(1)}_{9t} = {p\over{\sqrt{p^2 + q^2}}} B_{9t} ~= ~B^{(1)}_{98}, \cr
& \cr
B^{(2)}_{8t} = {q\over{\sqrt{p^2 + q^2}}} B_{8t},
~~B^{(2)}_{9t} = {q\over{\sqrt{p^2 + q^2}}} B_{9t} ~= ~B^{(2)}_{98}.
\label{pqdelocb}
\end{eqnarray}
Here $B_{8t}, B_{9t}, B_{98}$ are given by (\ref{10db}). The superscripts 
$1$ and $2$ indicate that these two forms are NS-NS and R-R
in nature respectively. 
Also, after $SL(2,Z)$ transformation, we  have the dilaton and axion as:
\begin{equation} 
\Phi_b^{(10)} = - 2 ~{\rm ln}\Big({(p^2 + q^2) \Delta ^{1\over 2}
\over{p^2 + q^2 \Delta}}\Big), ~~\chi = {pq(\Delta -1)\over{p^2 + q^2 
\Delta}}.
\label{pqdlocc}
\end{equation}
\subsection{Construction of Bound States}

Now, we perform, as before, a rotation on the above $(p,q)$ string solution
in the
$x^9-x^8$ plane. This is again given by
\begin{eqnarray}
dx^9 &=& {\rm cos} \phi ~d{\tilde x}^9 - {\rm sin}\phi ~d{\tilde x^8},
\nonumber\\
dx^8 &=& {\rm cos}\phi ~d{\tilde x}^8 + {\rm sin}\phi ~d{\tilde x}^9
\label{rot}.
\end{eqnarray}
Under this rotation, (\ref{pqdeloca}) - (\ref{pqdelocb}) become:
\begin{eqnarray}
ds^2 &=& {A\over \Delta}\Big( -(1 - \delta_\alpha )dt^2 
 +  [  {\rm sin}^2 \phi (1 + \delta_\alpha)
  + \gamma_\alpha {\rm sin}\phi {\rm cos}\phi + \Delta
   {\rm cos}^2\phi] (d\tilde x^9)^2 \nonumber\\
 &+&  [  {\rm cos}^2 \phi (1 + \delta_\alpha)
  - \gamma_\alpha {\rm sin}\phi {\rm cos}\phi + \Delta 
   {\rm sin}^2\phi] (d\tilde x^8)^2\nonumber\\
&+& [ \gamma_\alpha {\rm cos} 2\phi 
   + 2 (\delta_\alpha - \Delta +1) {\rm cos}\phi {\rm sin}\phi ] 
    d\tilde x^9 d\tilde x^8 
+ [ 2\delta_\alpha {\rm sin}\phi + \gamma_\alpha
 {\rm cos}\phi ] d\tilde x^9 dt\nonumber\\ 
&+& [ 2\delta_\alpha {\rm cos}\phi - \gamma_\alpha
 {\rm sin}\phi ] d\tilde x^8 dt + \Delta \sum_{i=1}^7 (dx^i)^2\Big),
\label{pqrota}
\end{eqnarray}
and 
\begin{eqnarray}
B^{(1)}_{\tilde 8 t} &=& {p \over{\Delta \sqrt{p^2 + q^2}}}[
{\rm cos}\phi (e^{-E}-1 + \delta_\alpha ) + {{\rm sin}\phi
\gamma_\alpha\over 2}],\cr
& \cr
B^{(1)}_{\tilde 9 t} &=& {p \over{\Delta \sqrt{p^2 + q^2}}}[
{\rm sin}\phi (e^{-E}-1 + \delta_\alpha ) - {{\rm cos}\phi
\gamma_\alpha\over 2}],\cr
& \cr
B^{(1)}_{\tilde 9 \tilde 8} &=& -{p \gamma_\alpha\over{2\Delta
\sqrt{p^2 + q^2}}}, \cr
& \cr
B^{(2)}_{\tilde 8 t} &=& {q \over{\Delta \sqrt{p^2 + q^2}}}[
{\rm cos}\phi (e^{-E}-1 + \delta_\alpha ) + {{\rm sin}\phi
\gamma_\alpha\over 2}],\cr
& \cr
B^{(2)}_{\tilde 9 t} &=& {q \over{\Delta \sqrt{p^2 + q^2}}}[
{\rm sin}\phi (e^{-E}-1 + \delta_\alpha ) - {{\rm cos}\phi
\gamma_\alpha\over 2}],\cr
& \cr
B^{(2)}_{\tilde 9 \tilde 8} &=& -{q \gamma_\alpha\over{2\Delta
\sqrt{p^2 + q^2}}}.
\label{pqrotb}
\end{eqnarray}
Here, we have defined 
\begin{equation}
\delta_\alpha = {\rm sinh}^2{\alpha\over 2}(e^{-E} -1),
~~\gamma_\alpha = {\rm sinh}\alpha (e^{-E} -1),
\nonumber
\end{equation}
Furthermore, the dilaton and 
axion remain same as (\ref{pqdlocc}). 

Our final step of the construction is to apply T-duality along $\tilde
x^9$ direction. The resulting solution would then correspond to 
a bound state in type IIA theory. The T-duality 
map between IIA and IIB theories is given explicitly in \cite{hull,
myers}. Applying this map, we 
end up with resulting configuration:
\begin{eqnarray}
d{\cal{S}}^2 &=& {1\over{ \Delta \sigma{\sqrt{p^2 + q^2}} {\sqrt{p^2 + q^2
   \Delta}}}}[p^2\{{\rm sin}\phi (e^{-E}  + \delta_\alpha)\nonumber\\
&&   ({\rm sin}\phi (e^{-E} -2 + \delta_\alpha) - {\rm cos}\phi
   \gamma_\alpha ) - \Delta {\rm cos}^2\phi (1 -
\delta_\alpha)\}\nonumber\\
  && - q^2\Delta\{ ({\rm sin}\phi + {\gamma_\alpha\over 2} {\rm
   cos}\phi)^2 
   + \Delta {\rm cos^2}\phi (1 - \delta_\alpha)\}] (dt)^2 \nonumber\\
&&+{1\over{ 2\Delta \sigma{\sqrt{p^2 + q^2}} {\sqrt{p^2 + q^2
  \Delta}}}}[\Delta\{-\gamma_\alpha^2 q^2 + 4 \delta_\alpha (
  p^2 + \Delta q^2)\}{\rm cos}\phi \nonumber\\
&&- 2 \gamma_\alpha
  \{ \delta_\alpha p^2 + e^{-E} p^2 + \Delta q^2\}{\rm sin}\phi]   
  dt d{\tilde x}^8\nonumber\\
&&+ {\Delta {\sqrt{p^2 + q^2}}\over{\sigma {\sqrt{p^2 + q^2 
  \Delta}}}}d\tilde x^9 d\tilde x^9\nonumber\\
&&- {p \over {\sigma{\sqrt{p^2 + q^2 \Delta}}}}[
  2 {\rm sin}\phi (e^{-E} -1 + \delta_\alpha) - \gamma_\alpha {\rm
  cos}\phi]d{\tilde x}^9 dt\nonumber\\
&&+ {1\over{\Delta \sigma {\sqrt{p^2 + q^2 \Delta}}{\sqrt{p^2 + q^2}}}}
  [ (\Delta + \delta_\alpha \Delta - {\gamma_\alpha^2\over 4})
  (p^2 + q^2 \Delta) \nonumber\\
&&+ p^2 \{{-\gamma_\alpha\over 2} {\rm cos}\phi
  + (e^{-E} -1 + \delta_\alpha) {\rm sin}\phi\}^2]
  d\tilde x^8 d\tilde x^8\nonumber\\
&& + {p \gamma_\alpha\over{\sigma{\sqrt{p^2 + q^2 \Delta}}}}d\tilde x^8
d\tilde x^9\nonumber\\
&& + {{\sqrt{p^2 + q^2 \Delta}}\over{\sqrt{p^2 + q^2}}}
\sum_{i=1}^{7}dx^i dx^i.
\label{iiametric}
\end{eqnarray}
The IIA dilaton is
\begin{equation}
e^{{\Phi}_a^{(10)}} = \Big({{{p^2 + q^2\Delta}}\over{{p^2 +
q^2}}}\Big)^{3\over 2}{1\over \sigma}.
\end{equation}
The two form components are given by
\begin{eqnarray}
{\cal{ B}}_{t \tilde 8} &=& -{{p\over {2\Delta\sigma\sqrt{p^2 +
  q^2}}}}
  [-\gamma_\alpha (e^{-E} + \delta_\alpha){\rm sin}\phi 
  +2 \Delta (-e^{-E} +1 - 
  \delta_\alpha){\rm cos}\phi],\nonumber\\
{\cal {B}}_{t \tilde 9} &=& {1\over \sigma}[\delta_\alpha {\rm
  sin}\phi +
  {\gamma_\alpha\over 2}{\rm cos}\phi],\nonumber\\
{\cal {B}}_{\tilde 8 \tilde 9} &=& {1\over \sigma}
[{\gamma_\alpha\over 2}{\rm cos} 2\phi + (\delta_\alpha- \Delta +1)
{\rm cos}\phi {\rm sin}\phi].
\label{iiab}
\end{eqnarray}
The nonzero components of 1-form and 3-form fields in IIA theory 
are given by
\begin{eqnarray}
{\cal {A}}_{t} &=& {q{\sqrt{p^2 + q^2}}\over{p^2 + q^2 \Delta}}
  [{\rm sin}\phi (e^{-E} -1 + \delta_\alpha ) - \gamma_\alpha{{\rm
  cos}\phi\over 2}],\nonumber\\
{\cal {A}}_{\tilde 8} &=& -{q \gamma_\alpha {\sqrt{p^2 + q^2}}\over
  {2(p^2
  + \Delta q^2)}} \nonumber\\
{\cal{A}}_{\tilde 9} &=& -{p q (\Delta -1)\over {p^2 +
  q^2\Delta}}\nonumber\\
{\cal{A}}_{t \tilde 8 \tilde 9} &=& {{q\over {2\sigma\Delta\sqrt{p^2
  +
  q^2}}}} [-\gamma_\alpha (e^{-E} + \delta_\alpha){\rm
  sin}\phi +2 \Delta (-e^{-E} +1 - 
  \delta_\alpha){\rm cos}\phi],
\label{iiaa}
\end{eqnarray}
where 
\begin{equation}
\sigma = {\rm sin}^2\phi ( 1 + \delta_\alpha) +
         \gamma_\alpha {\rm sin}\phi {\rm cos}\phi + 
         \Delta {\rm cos}^2\phi,
\end{equation}
and 
\begin{equation}
e^{-E} = 1 + {C \over r^5}. 
\end{equation}
We, once again, have multiple nonzero components of the 
NS-NS $2$-form as well as R-R $1$-form and $3$-form fields, in addition to 
having nonzero momenta coming from the off-diagonal components of the 
metric. We have explicitly checked that the above configuration 
reduces to known D-brane bound states in appropriate limits.
\subsection {Mass-Charge Relationship}

As in section-(4.2), we now verify that mass and charge associated 
with the above solution satisfy the expected mass-charge relation of
$1/2$ (non-threshold) BPS bound states. As in section-(4.2),we 
once again reduce the solution along both its spatial isometry 
directions. In the resulting theory in $D=8$, nonzero charges
are associated with gauge fields from the 
IIA field reductions. We have the non zero gauge fields which are of 
electric type, given as:
$A^1_t = {\cal{A}}_{t}$, 
$A^2_t = {\cal{A}}_{t\tilde 8 \tilde 9}$,
$A^3_t = -{\cal{B}}_{t \tilde 8}$, 
$A^4_t = {\cal{B}}_{t \tilde 9}$,
$A^5_t = g_{t \tilde 8}/g_{\tilde 8 \tilde 8}$, 
$A^6_t = g_{t \tilde 9}/g_{\tilde 9 \tilde 9}$ \cite{roy}. 
The charges can be read off from 
the leading order behavior of the above expressions, and are as
follows:
\begin{eqnarray}
Q_{1} &=& {{5 C q\over{\sqrt{p^2 + q^2}}}\times{{\rm cosh}{\alpha \over2}}
 ({\rm sin}{\phi}{\rm cosh}{\alpha\over2} 
- {\rm cos}{\phi}{\rm sinh}{\alpha\over2})},\cr
\cr
Q_{2} &=& -{{5 C q\over{\sqrt{p^2 + q^2}}}\times{{\rm cosh}{\alpha
\over2}}
({\rm cos}{\phi}{\rm cosh}{\alpha \over2} 
+ {\rm sin}{\phi}{\rm sinh}{\alpha \over2})},\cr
\cr
Q_{3} &=& - {{5 C p\over{\sqrt{p^2 + q^2}}}\times{{\rm cosh}{\alpha \over2}}
({\rm cos}{\phi}{\rm cosh}{
\alpha \over2} + {\rm sin}{\phi}{\rm sinh}{\alpha \over2})},\cr
\cr
Q_{4} &=& {5 C \times{{\rm sinh}{\alpha \over2}} ({\rm sin}{\phi}
{\rm sinh}{\alpha \over2} +
{\rm cos}{\phi}{\rm cosh}{\alpha \over2})},\cr
\cr
P_{1} &=& {5 C \times{{\rm sinh}{\alpha \over2}}({\rm cos}{\phi}{\rm
    sinh}{\alpha\over2} -
{\rm sin}{\phi}{\rm cosh}{\alpha \over2})},\cr
\cr
P_{2} &=& {{5 C p\over{\sqrt{p^2 + q^2}}}\times{{\rm cosh}{\alpha \over2}}
({\rm sin}{\phi}{\rm cosh}{\alpha \over2} - {\rm cos}{\phi}{\rm
  sinh}{\alpha \over2})},
\label{charges}
\end{eqnarray}
where $Q_i$ corresponds to the charge associated with
$A^i_t$, $(i=1,..,4)$ and the $P_i$, $(i=1,2)$ correspond to
the charges with respect to the gauge fields 
$A^5_t$ and $A^6_t$. They also correspond to the 
momenta along $\tilde{x}^8$ and $\tilde{x}^9$ in the original 
uncompactified theory. 
The ADM mass density of the bound state, that we have constructed 
is 
\begin{equation}
m^{(p, q)}_{(0,2)} = 5 C {\rm cosh}{\alpha}.
\label{massform}
\end{equation}
Therefore we have 
\begin{equation}
\left(m^{(p, q)}_{(0,2)}\right)^2 = {Q_{1}}^2 + {Q_{2}}^2 + 
{Q_{3}}^2 + {Q_{4}}^2 + {P_{1}}^2 + {P_{2}}^2.
\label{masscharge} 
\end{equation}
This relation implies that the bound state constructed above do satisfy
the BPS bound for the system.
As in section-4, one can also give a generalization 
of $Dp-D(p+2)$ bound states in a similar manner, as discussed above, 
by smearing 
directions $x^7$ etc. in the solution (\ref{iiametric} - \ref{iiaa}) 
and then by applying T-duality along these directions. We, however,
skip these details.   

It is interesting to note that we have been able to generate all six
gauge charges in $D=8$, using our procedure. As is known, these gauge
charges form a $(3, 2)$ representation of $SL(3, Z) \times SL(2, Z)$ 
$U$-duality symmetry in $D=8$. 
We can therefore rewrite the RHS of equation (\ref{masscharge})
in a $U$-duality invariant form by exciting general moduli as in 
\cite{schwarz,roy}. This is not surprising, as solution generating 
technique used by us is known to be equivalent to the one using 
$U$-duality transformations\cite{townsend,papa,cvetic}. 
In particular, in examples where we 
consider particle-like states in $D=8$ for writing down the 
mass-charge relations, the relevant $U$-duality group is 
$SL(3, Z)\times SL(2, Z)$. In type IIB examples, this $SL(2, Z)$ in 
$D=8$ originates from the group of constant coordinate transformations
along the two internal directions, whereas $SL(3, Z)$ is a
combination of the $D=10$ S-duality group together with the
$T$-duality along the $x^8$ and $x^9$. $(3, 2)$ multiplet of states
mentioned above are then generated by applying these transformations
to a seed solution in $D=8$, 
orginating from the (delocalized) F-string solution 
in $D=10$. All our transformations can also be mapped to 
appropriate ones lying inside the $U$-duality group.

\sect{\bf Conclusion}
In this paper, we have explicitly constructed nontrivial bound states
of D- branes starting with charged macroscopic strings. In particular, we 
were able to construct configurations in ten dimensions which carry 
$(F, D0, D2)$ charges as well as non-zero momenta. We also found that
these bound states are supersymmetric. We checked that all our 
solutions reduce to known bound states in appropriate limits.  We then
further generalized our results to $(Dp - D(p+2))$ bound states in 
IIA/B theories.

The bound states of D-branes,
when compactified to lower dimensions, often allow us to 
understand various properties of black hole including
Bekenstein-Hawking entropy in a
microscopic way \cite{ashokesen,wald,callan-malda}. 
It would be interesting to investigate as to what
new insight we gain from our configurations along this direction. A first
step may then be to reduce the solutions that 
we presented in sections-(3.1) and (4.1) to $D=8$ and $D=7$ 
respectively to interpret them as  charged  extremal black holes in
lower dimensions.
Various excitations of the higher dimensional bound states along
the compactified direction may then correspond to the required degrees
of freedom responsible for the entropy associated with the black hole. We
hope to persue along this direction in future.
It would also, perhaps, be interesting to understand the conformal field
theory
descriptions of these bound states. They are typically described by
boundary states of the underlying open string theory. Construction of
these boundary states often turned out to be very useful in the past,
see for example \cite{divecchia,lerda}.

\end{document}